\documentclass[fleqn,10pt]{wlscirep}
\usepackage[utf8]{inputenc}
\usepackage[T1]{fontenc}

\usepackage{caption}
\usepackage{subcaption}
\usepackage{tabularx}
\usepackage{booktabs}
\usepackage{graphicx}
\usepackage{ragged2e}
\usepackage{hyperref}

\usepackage{soul,color}

\title{Classification of Brain Tumours in MR Images using Deep Spatiospatial Models}

\author[1,2,3,*,+]{Soumick Chatterjee}
\author[4,+]{Faraz Ahmed Nizamani}
\author[2,3,5]{Andreas N\"urnberger}
\author[1,5,6,7]{Oliver Speck}

\affil[1]{Biomedical Magnetic Resonance, Otto von Guericke University Magdeburg, Germany}
\affil[2]{Data and Knowledge Engineering Group, Otto von Guericke University Magdeburg, Germany}
\affil[3]{Faculty of Computer Science, Otto von Guericke University Magdeburg, Germany}
\affil[4]{Institute for Medical Engineering, Otto von Guericke University Magdeburg, Germany}
\affil[5]{Center for Behavioral Brain Sciences, Magdeburg, Germany}
\affil[6]{German Center for Neurodegenerative Disease, Magdeburg, Germany}
\affil[7]{Leibniz Institute for Neurobiology, Magdeburg, Germany}

\affil[*]{soumick.chatterjee@ovgu.de}

\affil[+]{these authors contributed equally to this work}


\begin{abstract}
A brain tumour is a mass or cluster of abnormal cells in the brain, which has the possibility of becoming life-threatening because of its ability to invade neighbouring tissues and also form metastases. An accurate diagnosis is essential for successful treatment planning, and magnetic resonance imaging is the principal imaging modality for diagnosing brain tumours and their extent. Deep Learning methods in computer vision applications have shown significant improvement in recent years, most of which can be credited to the fact that a sizeable amount of data is available to train models, and the improvements in the model architectures yield better approximations in a supervised setting. Classifying tumours using such deep learning methods has made significant progress with the availability of open datasets with reliable annotations. Typically those methods are either 3D models, which use 3D volumetric MRIs or even 2D models considering each slice separately. However, by treating one spatial dimension separately or by considering the slices as a sequence of images over time, spatiotemporal models can be employed as "spatiospatial" models for this task. These models have the capabilities of learning specific spatial and temporal relationships while reducing computational costs. This paper uses two spatiotemporal models, ResNet (2+1)D and ResNet Mixed Convolution, to classify different types of brain tumours. It was observed that both these models performed superior to the pure 3D convolutional model, ResNet18. Furthermore, it was also observed that pre-training the models on a different, even unrelated dataset before training them for the task of tumour classification improves the performance. Finally, Pre-trained ResNet Mixed Convolution was observed to be the best model in these experiments, achieving a macro F1-score of 0.9345 and a test accuracy of 96.98\%, while at the same time being the model with the least computational cost. 
\end{abstract}
\begin{document}

\flushbottom
\maketitle
%
%
\thispagestyle{empty}

\section{Introduction}
A brain tumour is the growth of abnormal cells in the brain. Brain tumours are classified based on their speed of growth and the likeness of them growing back after treatment. They are mainly divided into two overall categories: malignant and benign. Benign tumours are not cancerous, they grow slowly and are less likely to return after treatment. Malignant tumours, on the other hand, are essentially made up of cancer cells, they have the ability to invade the tissues locally, or they can spread to different parts of the body, a process called metastasise \cite{fritz2000international}. \textbf{Glioma tumours} are the result of glial cell mutations resulting in malignancy of normal cells. They are the most common types of Astrocytomas (tumour of the brain or spinal cord), account for $30\%$ of all brain and central nervous system tumours, and $80\%$ of all malignant tumours~\cite{goodenberger2012genetics}. The phenotypical makeup of glioma tumours can consist of Astrocytomas, Oligodendrogliomas, or Ependymomas. Each of these tumours behaves differently, and World Health Organisation (WHO) uses the following grading-based method to categorise each tumour based upon its aggressiveness:

\begin{itemize}
\item \textbf{Grade I} tumours are generally benign tumours, which means they are mostly curable, and they are commonly found in children.

\item \textbf{Grade II} includes three types of tumours: Astrocytomas, Oligodendrogliomas, and Oligoastrocytoma - which is a mix of both~\cite{claus2015survival}. They are common in adults. Eventually, all low-grade gliomas can progress to high-grade tumours~\cite{claus2015survival}.

\item \textbf{Grade III} tumour can include Anaplastic Astrocytomas, Anaplastic Oligodendrogliomas or Anaplastic Oligoastrocytoma. They are more aggressive and infiltrating than grade II.

\item \textbf{Grade IV} glioma, also called \textbf{Glioblastoma Multiforme (GBM)}, is the most aggressive tumour in the WHO category. 

\end{itemize}

\begin{figure}[h!]

\begin{center}
	\centering
		\includegraphics[width=0.5\textwidth]{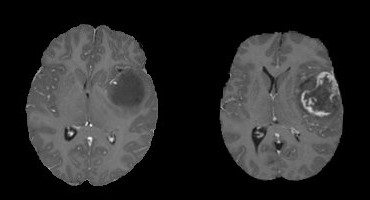}
	
	\caption{An example MRI of Low-grade glioma (LGG, on the left) and High-grade glioma (HGG, on the right), Source: BraTS 2019}
	\label{fig:lgghgg}

\end{center}
\end{figure}

In general, grades I and II gliomas are considered low-grade gliomas (LGG), while grades III and IV are known as high-grade glioma (HGG). The LGG are benign tumours, and they can be excised using surgical resection. In contrast, HGGs are malignant tumours that are hard to excise by surgical methods because of their extent of nearby tissue invasion. Fig.~\ref{fig:lgghgg} shows an example MRI of LGG and HGG.

\begin{figure}[h!]

\begin{center}
	\centering
		\includegraphics[width=0.7\textwidth]{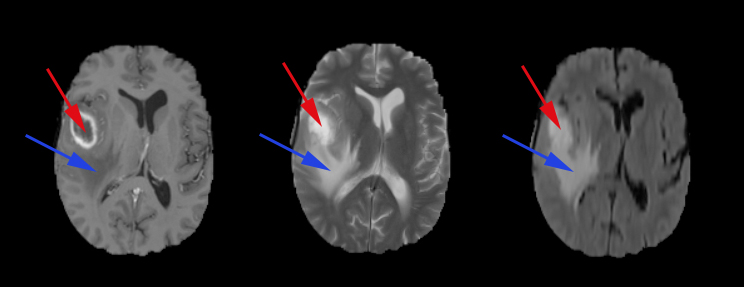}
	
	\caption{High-grade glioma structure on T1ce, T2 and FLAIR contrast images (from left to right), (\textcolor{red}{•}) Necrotic core, (\textcolor{blue}{•}) Perifocal oedema, Source: BraTS 2019}
	\label{fig:gbm}

\end{center}

\end{figure}

A Glioblastoma Multiforme (GBM) typically has the following types of tissues (shown in Fig.~\ref{fig:gbm}):

\begin{itemize}
  \item \textbf{The Tumour Core}:  This is the region of the tumour that has the malignant cells that are actively proliferating.
  \item \textbf{Necrosis}: The necrotic region is the important distinguishing factor between low-grade gliomas and GBM~\cite{raza2002necrosis}. This is the region where the cells/tissue are dying, or they are dead. 
  \item \textbf{Perifocal oedema}: The swelling of the brain is caused by fluid build-up around the tumour core, which increases the intracranial pressure; perifocal oedema is caused by the changes in glial cell distribution~\cite{engelhorn2009cellular}. 
\end{itemize}

The prognosis of a brain tumour depends on many factors, such as the tumour's location, the histological subtype of the tumour, and the tumour margins. In many cases, the tumour reoccurs and progresses to grade IV even after treatment~\cite{claus2015survival}. Modern imaging methods such as MRI can be used for multiple diagnostic purposes; they can be used to identify the tumour location - which is used for investigating tumour progression and surgical pre-planning. MR imaging is also used to study the anatomy of the lesion, physiology, and metabolic activity along with its haemodynamics. Therefore MR imaging remains the primary diagnostic modality for brain tumours.

Detection of cancer, specifically an earlier detection, holds the potential to make a difference in treatment. Earlier detection is vital because lesions in earlier stages are more likely curable; therefore, if intervened early on, this can make the difference between life and death. Deep learning methods can help automate the process of detecting and classifying brain lesions - they can also reduce the radiologists' burden of reading many images by prioritising only malignant lesions. This will eventually improve the overall efficiency, and it can reduce diagnostic errors~\cite{menze2014multimodal}. Recent studies have shown that deep learning methods in the field of radiology have already achieved comparable and super-human performance for some pathologies~\cite{rajpurkar2018deep}. 

\subsection{Related Work}
Various deep learning based methods have been proposed in recent times to classify brain tumours. Mzoughi et al.~\cite{mzoughi2020deep} proposed an approach using volumetric CNNs to classify high-grade glioma and low-grade glioma using T1 contrast-enhanced images. Another similar work on glioma classification based on grading was done by Pei et al.~\cite{pei2019brain}, where they first segmented the tumour and then classified the tumour between HGG and LGG. Most of the literature on glioma tumour classification and grading used one single MR contrast image at a time, but Ge et al.~\cite{ge2018deep} used a fusion framework that uses T1 contrast-enhanced, T2, and FLAIR images simultaneously for classifying the tumour. Ouerghi et al.~\cite{ouerghi2021glioma} used a novel fusion method for the inclusion of multiple MRI contrasts, first, the T1 images are transformed by non-subsampled shearlet transform (NSST) into low frequency (LF) and high frequency (HF) subimages, essentially separating principle information in the source image from edge information, then the images are fused by predefined rules to include the coefficients, resulting in fusion of T1 and T2 or FLAIR images. Most of the literature only classifies between the different grades of tumour and does not consider healthy brains as an additional class.

\subsection{Technical Background}
ResNet or residual network, proposed by He et al.~\cite{he2016deep}, has shown to be one of the most efficient network architectures for image recognition tasks, dealing with problems of deep networks, e.g. vanishing gradients. This paper introduced residual-link, the identity mappings, which are "skipped connections", whose outputs are added to the outputs of the rest of the stacked layers.  These identity connections do not add any complexity to the network while improving the training process. The spatiotemporal models introduced by Tran et al.~\cite{tran2018closer} for action recognition are fundamentally 3D Convolutional Neural Networks based on ResNet. There are two spatial dimensions and one temporal dimension in video data, making the data three dimensional. For handling such data (e.g. action recognition task), using a network with 3D convolution layers is an obvious choice. Tran et al.~\cite{tran2018closer} introduced two variants of spatiotemporal models: ResNet (2+1)D and ResNet Mixed Convolution. The ResNet(2+1)D model consists of 2D and 1D convolutions, where the 2D convolutions are used spatially while the 1D convolutions are reserved for the temporal element. This gives an advantage of increased non-linearity by using non-linear rectification, which allows this kind of mixed model to be more "learnable" than conventional full 3D models. On the other hand, the ResNet Mixed Convolution model is constructed as a mixture of 2D and 3D Convolution operations. The initial layers of the model are made of 3D convolution operations, while the later layers consist of 2D convolutions. The rationale behind using this type of configuration is that the motion-modelling occurs mostly at the initial layers, and applying 3D convolution there encapsulates action better. 

Apart from trying to improve the network architecture, one frequently used technique to improve the performance of the same architecture is transfer  learning~\cite{torrey2010transfer}. This is a technique for re-purposing a model for another task that is different from the task the model was originally trained for performing. Typically, the model parameters are initialised randomly before starting the training. However, in the case of transfer learning, model parameters learned from task one are used as the starting point (called pre-training), instead of random values, for training the model for task two. Pre-training has shown to be an effective method to improve the initial training process, eventually achieving better accuracy~\cite{zhuang2020comprehensive,sarasaen2021fine}. 

\subsection{Contribution}
Spatiotemporal models are typically used for video classification tasks, which are three dimensional in nature. Their potential in classifying 3D volumetric images like MRI, considering them as "spatiospatial" models, has not been explored yet.
This explores the possibility of applying spatiotemporal models (ResNet(2+1)D and ResNet Mixed Convolution) as "spatiospatial" models by treating one dimension (slice dimension) differently than the other two spatial dimensions of the 3D volumetric images. "Spatiospatial" were employed to classify brain tumours of the different types of gliomas based on their grading as well as healthy brains from 3D volumetric MR Images using a single MR contrast, and compare their performances against a pure 3D convolutional model (ResNet3D). Furthermore, the models are to be compared with and without pre-training - to judge the usability of transfer learning for this task.

\section{Methodology}
This section explains the network models used in this research, implementation details, pre-training and training methods, data augmentation techniques, dataset information, data pre-processing steps, and finally, the evaluation metrics.

\subsection{Network Models}
Spatiotemporal models are mainly used for video-related tasks, where there are two spatial and one temporal dimension. These models deal with the spatial and temporal dimensions differently, unlike pure 3D convolution-based models. There is no temporal component in 3D volumetric image classification tasks; hence, using a 3D convolution-based model is a frequent choice. At times, they are divided into 2D slices, and 2D convolution-based models are applied to them. For the task of tumour classification, the rationale for using 3D filters is grounded in the morphological heterogeneity of gliomas~\cite{pallud2012quantitative}, it is to make the convolution kernels invariant to tissue discrimination in all dimensions, learning more complex features spanning voxels, while 2D convolution filters will capture the spatial representation within the slices. Spatiotemporal models combine two different types of convolution into one model while having the possibility of reducing the complexity of the model or of incorporating more non-linearity. These advantages might be possible to exploit while working with volumetric data by considering the spatiotemporal models as "spatiospatial" models - the motivation behind using such models for a tumour classification task. In this paper, the slice-dimension is treated as the pseudo-temporal dimension of spatiotemporal models, and in-plane dimensions are treated as the spatial dimensions. The spatiotemporal models used here as spatiospatial models are based on the work of Tran et al.~\cite{tran2018closer}. 

Two different spatiospatial models are explored here: ResNet (2+1)D and ResNet Mixed Convolution. Their performances are compared against ResNet3D, which is a pure 3D convolution-based model. 

\begin{figure}[h!]
\begin{center}
    \centering

    	\includegraphics[width=0.8\textwidth]{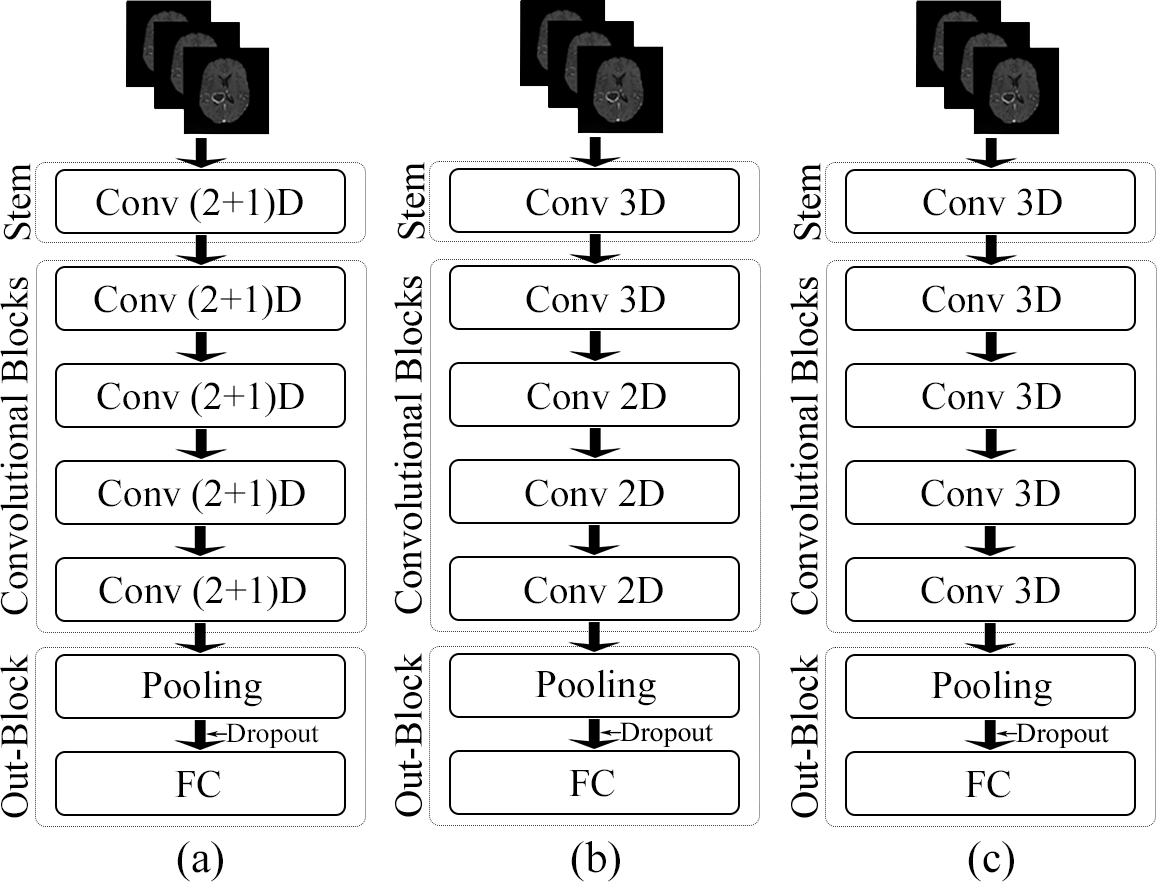}
	
	\caption{Schematic representations of the network architectures. (a) ResNet (2+1)D, (b) ResNet Mixed Convolution, and (c) ResNet 3D}
	\label{fig:nets}

\end{center}

\end{figure}

\subsubsection{ResNet (2+1)D}
ResNet (2+1)D uses a combination of 2D convolution followed by 1D convolution instead of a single 3D convolution. The benefit of using this configuration is that it allows an added non-linear activation unit between the two convolutions, as in comparison to using a single 3D Convolution~\cite{tran2018closer}. This then results in an overall increase of ReLU units in the network, giving the model the ability to learn even more complex functions. The ResNet(2+1)D uses a stem that contains a 2D convolution with a kernel size of seven and a stride of two, accepting one channel as an input and providing 45 channels as output; followed by a 1D convolution with a kernel size of three and a stride of one, providing 64 channels as the final output. Next, there are four convolutional blocks; each of them contains two sets of basic residual blocks. Each residual block contains one 2D convolution with a kernel size of three and a stride of one, followed by a 1D convolution with a  kernel size of three and a stride of one. Each convolutional layer in the model (both 2D and 1D) is followed by a 3D batch normalisation layer and a ReLU activation function. The residual blocks inside the convolutional blocks, except for the first convolutional block, are separated by a pair of 3D convolution layers with a kernel size of one and a stride of two - to downsample the input by half. The 2D convolutions are applied in-plane, and the 1D convolutions are applied on the slice dimension. After the final convolutional block, an adaptive average pooling layer has been added, with an output size of one for all three dimensions. After the pooling layer, a dropout layer followed by a fully connected layer with $n$ output neurons for $n$ classes were added to obtain the final output. Fig.~\ref{fig:nets}(a) portrays the schematic diagram of the ResNet (2+1)D architecture. 

\subsubsection{ResNet Mixed Convolution}
\label{sec:mc}
ResNet Mixed Convolution uses a combination of 2D and 3D Convolutions. The stem of this model contains a 3D convolution layer with a kernel size of (3,7,7), a stride of (1,2,2), and padding of (1,3,3) - where the first dimension is the slice dimension and the other two dimensions are the in-plane dimensions, and accepts a single channel as input while providing 64 channels as output. After the stem, there is one 3D convolution block, followed by three 2D convolution blocks. All the convolution layers (both 3D and 2D) have a kernel size of three and a stride of one, identical for all dimensions. Each of these convolution blocks contains a pair of residual blocks, each of which contains a pair of convolution layers. Similar to ResNet (2+1)D, the residual blocks inside the convolutional blocks, except for the first convolutional block, are separated by a pair of 3D convolution layers with a kernel size of one and a stride of two - to downsample the input by half. Each convolutional layer in the model (both 3D and 2D) is followed by a 3D batch normalisation layer and a ReLU activation function. The motivation behind using both modes of convolution in 2D and 3D is that the 3D filters can learn the spatial features of the tumour in 3D space while 2D can learn representation within each 2D slice. After the convolutional blocks, the final pooling, dropout, and fully connected layers are identical to the ResNet (2+1)D architecture. Fig.~\ref{fig:nets}(b) shows the schematic representation of this model.

\subsubsection{ResNet3D}
The performance of the spatiospatial models are compared against a pure 3D ResNet model, schematic diagram shown in Fig.~\ref{fig:nets}(c). The architecture of the ResNet3D model used here is almost identical to the architecture of ResNet Mixed Convolution (Section~\ref{sec:mc}), except for the fact that this model uses only 3D convolutions. The stem of these models are identical, the only difference being that this model uses four 3D convolution blocks, unlike ResNet Mixed Convolution, which uses one 3D convolution block, followed by three 2D convolution blocks. This configuration of ResNet3D architecture results in a 3D ResNet18 model.

\subsubsection{Summary and Comparison}
The general structure of the network models can be divided into the following: input goes to the stem, then there are four convolutional blocks, followed by the output block - which contains an adaptive pooling layer, followed by a dropout layer, and finally a fully connected layer. ResNet Mixed Convolution and ResNet 3D have the same stem, including a 3D convolutional layer with a kernel size of (3,7,7), followed by a batch normalisation layer and a ReLU. ResNet (2+1)D uses a different stem: a 2D convolution layer with a kernel size of seven, then a 1D convolution with a kernel size of three - splitting the 3D convolution (3,7,7) used by the other models into a pair of 2D and 1D convolution: (7,7) and (3). Both 2D and 1D convolution inside this stem is followed by a batch normalisation layer and ReLU pair. The convolutional blocks in the ResNet3D and ResNet Mixed Convolution architectures follow the same architecture: two residual blocks consisting of two sub-blocks consisting of a 3D convolution with a kernel size of three, followed by batch normalisation layer and a ReLU. On the other hand, the first convolutional block of the ResNet (2+1)D architecture uses a pair of 2D and 1D convolutions with the kernel size of three instead of the 3D convolutional layers used by the other models. The rest of the architecture is the same. It is noteworthy that this model has more non-linearity because the 3D convolutions are split into a pair of 2D and 1D convolutions; additional pair of batch normalisation and ReLU could have been used between the 2D 1D convolution. There is one difference between the first convolutional block and the other three blocks (applicable for all three models): the second, third and fourth convolutional blocks included a downsampling pair, which consisted of a 3D convolutional layer with a kennel size of one and a stride of two, followed by a batch normalisation layer. This was not present in the first convolutional block. The convolution blocks of each of all three models double the input features by two (number of input features to the first block: 64, number of output features of the fourth (and final) block: 512). All of these models end with an adaptive average pooling layer, which forces the output to have a shape of 1x1x1, with 512 different features. A dropout with a probability of 0.3 is then applied to introduce regularisation to prevent over-fitting before supplying them to a fully connected linear layer that generates $n$ classes as output. The width and depth of these models are comparable, but they differ in terms of the number of trainable parameters depending upon the type of convolution used, as shown in Tab.~\ref{tab:params}. It is noteworthy that the less the number of trainable parameters - the less the computational costs. A model with a lesser number of parameters would require lesser memory for computation (GPU and RAM), and also the complexity of the model is lesser - reducing the overall computational costs for both training and inference. Moreover, a lesser number of trainable parameters would also reduce the risk of overfitting.

\begin{table}[h!]
\centering
\caption{Total Number of Trainable parameters for each model}
\label{tab:params}
\renewcommand*\arraystretch{1.5}
\begin{tabular}[t]{lc}
\hline
\toprule
 \textbf{Model} & No of Parameters   \\
\toprule
\textbf{ResNet3D} & 33,150,522\\
\textbf{ResNet (2+1)D} & 31,297,254\\
\textbf{ResNet Mixed Convolution} & 11,472,963\\

\toprule
\end{tabular}

\end{table}

\subsection{Implementation and Training}
The models were implemented using PyTorch~\cite{paszke2019pytorch}, by modifying the Torchvision models~\cite{torchvismodels} and were trained with a batch-size of $1$ using an Nvidia RTX 4000 GPU, which has a memory of $8$GB. Models were compared with and without pre-training. Models with pre-training were pre-trained on Kinetics-400~\cite{kinetics400dataset}, except for the stems and fully connected layers. Images from the Kinetics dataset contain three channels (RGB Images), whereas the 3D volumetric MRIs have only one channel. Therefore, the stem trained on the Kinetics dataset could not be used and was initialised randomly. Similarly, for the fully connected layer, Kinetics-400 has 400 output classes, whereas the task at hand has three classes (LGG, HGG and Healthy) - hence, this layer was also initialised with random weights. 

Trainings were performed using mixed-precision~\cite{micikevicius2017mixed} with the help of Nvidia's Apex library~\cite{apex}. The loss was calculated using the weighted cross-entropy loss function to minimise the under-representation of classes with fewer samples during training and was optimised using the Adam optimiser with a learning rate of 1e-5 and weight decay coefficient $\lambda$=1e-3. The code of this research is publicly available on GitHub: \url{https://github.com/farazahmeds/Classification-of-brain-tumor-using-Spatiotemporal-models}.

\subsubsection{Weighted Cross-entropy Loss}
\label{sec:loss}
The normalised weight value for each class ($W_c$) is calculated using:
\begin{equation}
W_c = \left[1 - \left( \frac {samples_c}{\Sigma{samples_t}}\right)\right]
\end{equation}
where $samples_c$ is the number of samples from class c and $samples_t$ are the total number of samples from all classes. 
The normalised weight values from this equation is then used to scale cross-entropy loss of the respective class loss:
\begin{equation}
loss_c =W_c\left[-x_{c} \log (P(c))\right]
\end{equation}
Where $x_{c}$ is the true distribution and P(c) is the estimate distribution for class c.
The total cross-entropy loss then is the sum of individual class losses.
\begin{equation}
Loss_{total} =loss_{c_1}+loss_{c_2}+loss_{c_2}+...+loss_{c_n}
\end{equation}

\subsection{Data Augmentation}
Different data augmentation techniques were applied to the dataset before training the models, and for that purpose, TorchIO~\cite{perez2021torchio} was used. Initial experiments were performed using different amounts of augmentation and can be categorised as light and heavy augmentation, where light augmentation included only random affine (scale 0.9-1.2, degrees 10) and random flip (L-R, probability 0.25); on the other hand, heavy augmentation included the ones from light augmentation together with elastic deformation and random k-space transformations (motion, spike, and ghosting). It was observed that the training of the network with heavily augmented data not only performed poorly in terms of final accuracy, but the loss took a much longer time to converge. Therefore, only light augmentation was used throughout this research.

\subsection{Dataset}
\label{sec:dataset}
Two different datasets were used in this work - the pathological brain images were obtained from the Brain Tumour Segmentation (BraTS) 2019 dataset, which includes images with four different MR contrasts (T1, T1 contrast-enhanced, T2 and FLAIR)~\cite{menze2014multimodal, bakas2017advancing, bakas2018identifying}; and non-pathological images were collected from the IXI Dataset~\cite{ixidataset}. Among the available four types of MRIs, T1 contrast-enhanced (T1ce) is the most commonly used contrast while performing single-contrast tumour classification~\cite{yang2018glioma,mzoughi2020deep}. Hence in this research, T1ce images of 332 subjects were used from the BRaTS dataset: 259 volumes of Glioblastoma Multiforme (high-grade glioma, HGG), and 73 volumes of low-grade glioma (LGG). 259 T1 weighted volumes were chosen randomly from the IXI dataset as healthy samples to have the same number of subjects as HGG. The final combined dataset was then randomly divided into 3-folds of training and testing split with a ratio of 7:3.

\subsection{Data Pre-processing}
The IXI images were pre-processed first by using the brain extraction tool (BET2) of FSL~\cite{smith2004advances,jenkinson2012smith}. This was done to keep the input data uniform throughout, as the BraTS images are already skull stripped. Moreover, the intensity values of all the volumes from the combined datasets were normalised by scaling intensities to [0.5,99.5] percentile, as used by Isensee et al~\cite{isensee2018nnu}. Finally, the volumes were re-sampled to the same voxel-resolution of 2mm isotropic. 

\subsection{Evaluation Metrics}
The performance of the models was compared using precision, recall, F1 score, specificity, and testing accuracy. Furthermore, a confusion matrix was used to show class-wise accuracy. 


\section{Results}
\begin{figure}[h!]
     \centering
     \begin{subfigure}[b]{0.49\textwidth}
         \centering
         \includegraphics[width=\textwidth]{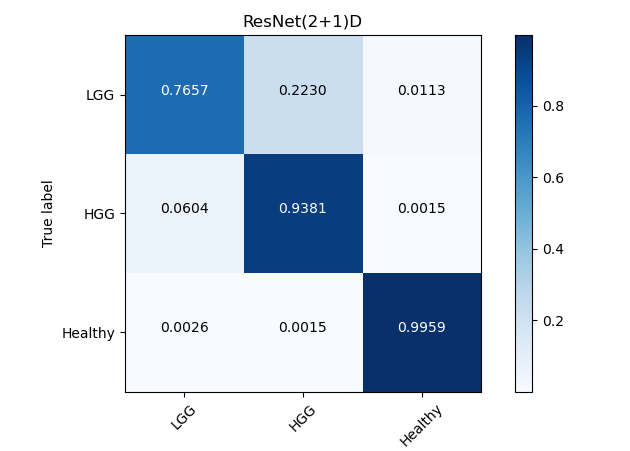}
         \caption{Without pre-training}
     \end{subfigure}
     \hfill
     \begin{subfigure}[b]{0.49\textwidth}
         \centering
         \includegraphics[width=\textwidth]{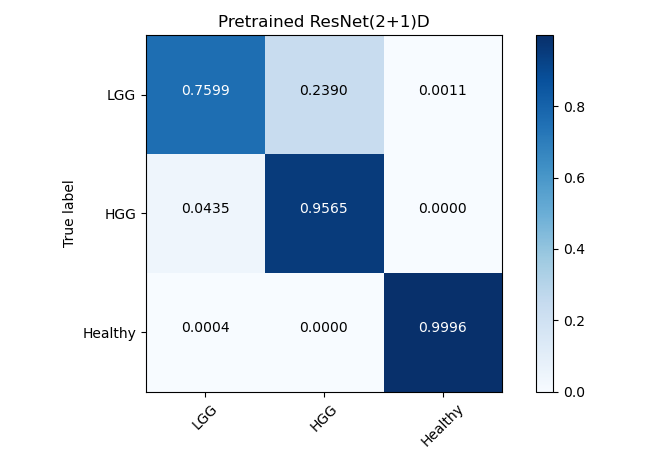}
         \caption{With pre-training}
     \end{subfigure}
\caption{Confusion Matrix for 3-fold cross-validation on Pre-trained ResNet(2+1)D}
\label{fig:conf_resnet2p1d}
\end{figure}

\begin{figure}[h!]
     \centering
     \begin{subfigure}[b]{0.49\textwidth}
         \centering
         \includegraphics[width=\textwidth]{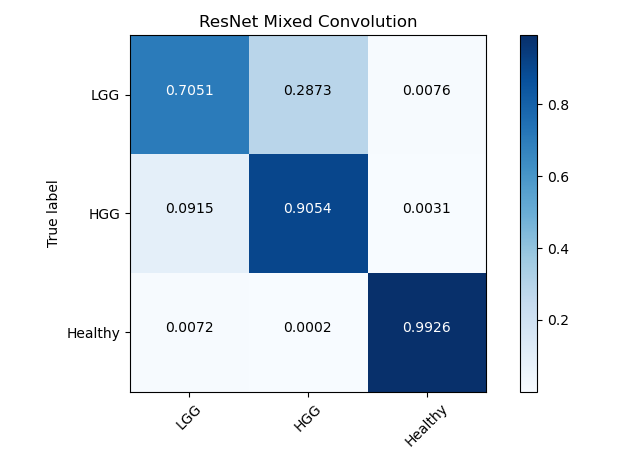}
         \caption{Without pre-training}
     \end{subfigure}
     \hfill
     \begin{subfigure}[b]{0.49\textwidth}
         \centering
         \includegraphics[width=\textwidth]{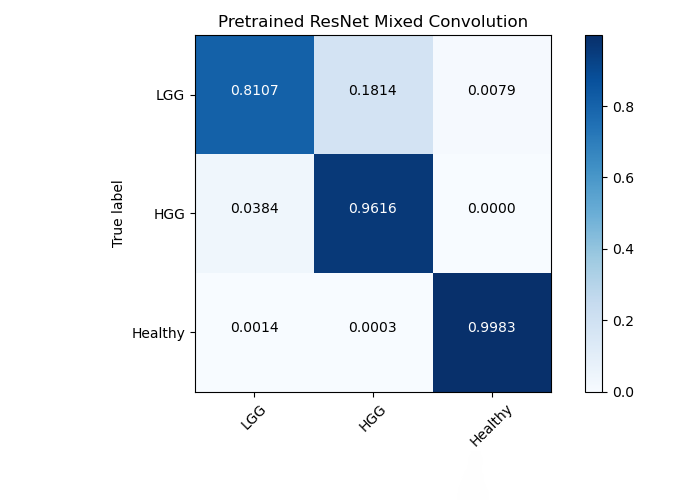}
         \caption{With pre-training}
     \end{subfigure}
\caption{Confusion Matrix for 3-fold cross-validation on ResNet Mixed Convolution}
\label{fig:conf_resnetmc}
\end{figure}

\begin{figure}[h!]
     \centering
     \begin{subfigure}[b]{0.49\textwidth}
         \centering
         \includegraphics[width=\textwidth]{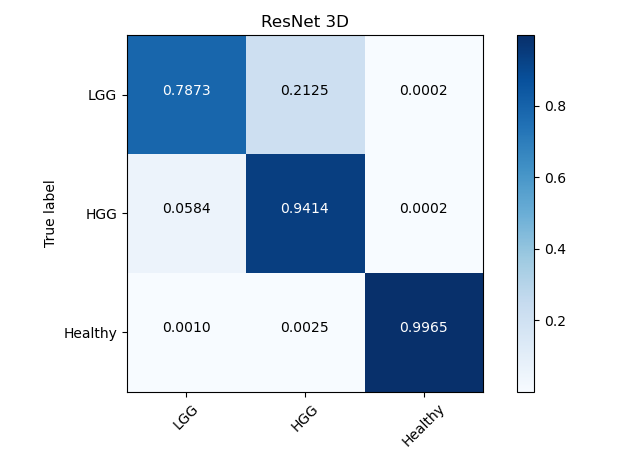}
         \caption{Without pre-training}
     \end{subfigure}
     \hfill
     \begin{subfigure}[b]{0.49\textwidth}
         \centering
         \includegraphics[width=\textwidth]{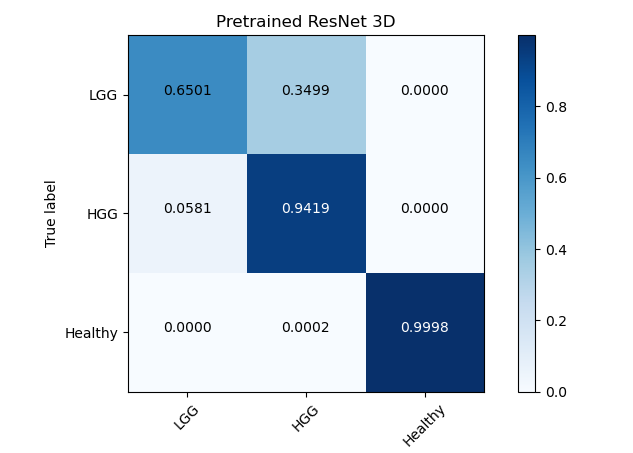}
         \caption{With pre-training}
     \end{subfigure}
\caption{Confusion Matrix for 3-fold cross-validation on ResNet3D18}
\label{fig:conf_resnet3d}
\end{figure}

The performance of the models were compared with and without pre-training. Figures \ref{fig:conf_resnet2p1d}, \ref{fig:conf_resnetmc}, and \ref{fig:conf_resnet3d} show the average accuracy over 3-fold cross validation using confusion metrics, for ResNet (2+1)D, ResNet Mixed Convolution, and ResNet 3D, respectively. 

\begin{figure}[h!]
\begin{center}
    \centering

    	\includegraphics[width=\textwidth]{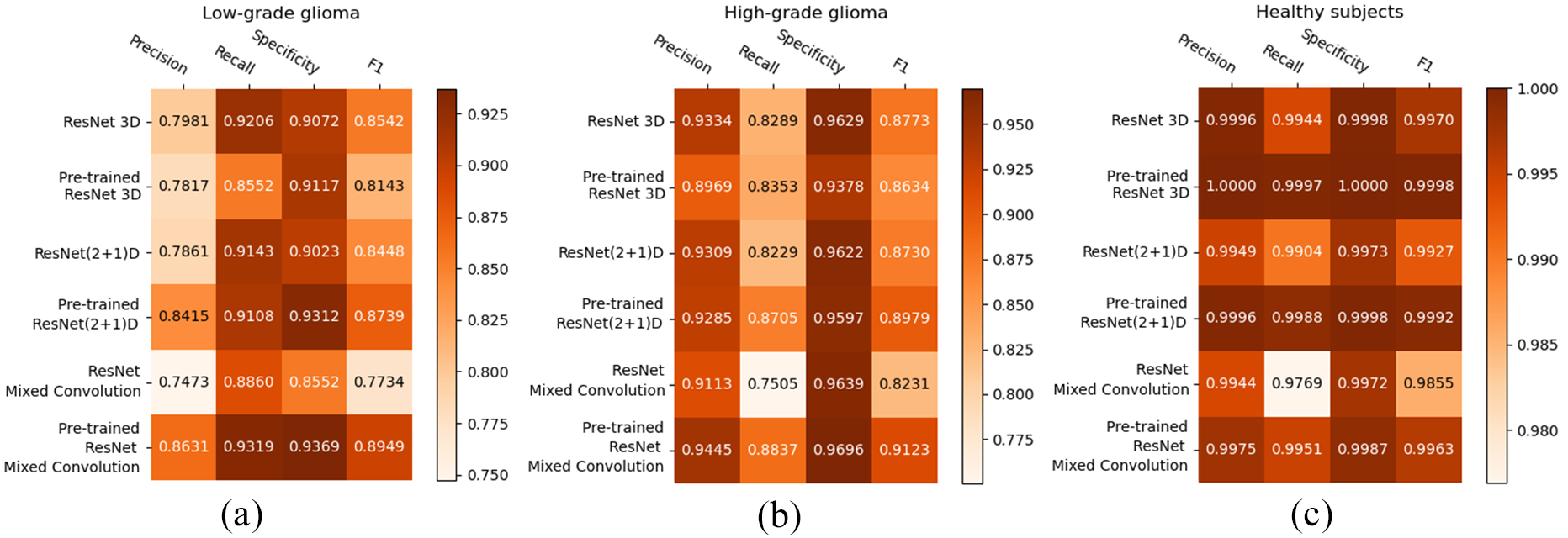}
	
	\caption{Heatmaps showing the class-wise performance of the classifiers, compared using Precision, Recall, Specificity, and F1-score: (a) LGG, (b) HGG, and (c) Healthy}
	\label{fig:classifierperform_diseasewise}

\end{center}

\end{figure}

Fig.~\ref{fig:classifierperform_diseasewise} shows the class-wise performance of the different models, both with and without pre-training, using precision, recall, specificity, and F1-score.

\subsection{Comparison of the models}
The mean F1-score over 3-fold cross-validation was used as the metric to compare the performance of the different models. Tables \ref{tab:f1_lgg}, \ref{tab:f1_hgg}, and \ref{tab:f1_healthy} show the results of the different models for the classes LGG, HGG, and Heathy, respectively; and finally Table~\ref{tab:f1_consolidated} shows the consolidated scores.

\begin{table}[h!]
\centering

\renewcommand*\arraystretch{1.5}
\begin{tabular}{lc}
\hline
    \toprule
         \multicolumn{2}{c}{\textbf{Low-grade glioma}} \\
\toprule
  \textbf{Model} & \textbf{mean F1 score}  \\
\toprule
ResNet 3D & 0.8542 ± 0.049  \\
\hline
Pre-trained ResNet 3D & 0.8143 ± 0.048 \\
\hline
ResNet(2+1)D & 0.8448 ± 0.019 \\
\hline
Pre-trained ResNet(2+1)D  & 0.8739 ± 0.041 \\
\hline
ResNet Mixed Convolution & 0.7734 ± 0.031 \\
\hline
\textbf{Pre-trained ResNet Mixed Convolution}\textcolor{orange}{*} & \textbf{0.8949 ± 0.033} \\

\toprule

\end{tabular}

\caption{Low-grade glioma model comparison (\textcolor{orange}{*} denotes the overall winning model)}
\label{tab:f1_lgg}
\end{table}%

For low-grade glioma (LGG), ResNet Mixed Convolution with pre-training achieved the highest F1 score of 0.8949 with a standard deviation of 0.033. The pre-trained ResNet(2+1)D is not far behind, with 0.8739 ± 0.033. 

\begin{table}[h!]
\centering

\renewcommand*\arraystretch{1.5}
\begin{tabular}{lc}
\hline
    \toprule
         \multicolumn{2}{c}{\textbf{High-grade glioma}} \\
\toprule
  \textbf{Model} & \textbf{mean F1 score}  \\
\toprule
ResNet 3D & 0.8773 ± 0.034 \\
\hline
Pre-trained ResNet 3D & 0.8634 ± 0.042  \\
\hline
ResNet(2+1)D & 0.8730 ± 0.022 \\
\hline
Pre-trained ResNet(2+1)D  & 0.8979 ± 0.032 \\
\hline
ResNet Mixed Convolution & 0.8231 ± 0.027  \\
\hline
\textbf{Pre-trained ResNet Mixed Convolution}\textcolor{orange}{*} & \textbf{0.9123 ± 0.029} \\

\toprule

\end{tabular}

\caption{High-grade glioma model comparison (\textcolor{orange}{*} denotes the overall winning model)}
\label{tab:f1_hgg}
\end{table}%

For the high-grade glioma (HGG) class, the highest F1 was achieved by the pre-trained ResNet Mixed Convolution model, with an F1 score of 0.9123 ± 0.029. This is higher than the best model's F1 score for the class LGG. This can be expected because of the class imbalance between LGG and HGG. As with low-grade glioma, the second-best model for HGG is also the Pre-trained ResNet(2+1)D with the F1 score of 0.8979 ± 0.032.

\begin{table}[h!]
\centering

\renewcommand*\arraystretch{1.5}
\begin{tabular}{lc}
\hline
    \toprule
         \multicolumn{2}{c}{\textbf{Healthy brain}} \\
\toprule
  \textbf{Model} & \textbf{mean F1 score}  \\
\toprule
ResNet 3D & 0.9970 ± 0.005 \\
\hline
\textbf{Pre-trained ResNet 3D} & \textbf{0.9998 ± 0.0002} \\
\hline
ResNet(2+1)D & 0.9927 ± 0.009 \\
\hline
Pre-trained ResNet(2+1)D  & 0.9992 ± 0.001 \\
\hline
ResNet Mixed Convolution &  0.9855 ± 0.004\\
\hline
Pre-trained ResNet Mixed Convolution\textcolor{orange}{*} & 0.9963 ± 0.002 \\

\toprule

\end{tabular}

\caption{Healthy brain model comparison (\textcolor{orange}{*} denotes the overall winning model)}
\label{tab:f1_healthy}
\end{table}%

The healthy brain class achieved the highest F1 score of 0.9998 ± 0.0002, with the pre-trained ResNet 3D model, which can be expected because of the complete absence of any lesion in the MR images making it far less challenging for the model to learn and distinguish it from the brain MRIs with pathology. Even though the pre-trained ResNet 3D model achieved the highest mean F1 score, all pre-trained models achieved similar F1 scores, i.e. all the mean scores are more than 0.9960 - making it difficult to choose a clear winner. 

\begin{table}[h!]
\centering

\renewcommand*\arraystretch{1.5}
\begin{tabular}{lcc}
\hline
    \toprule
         \multicolumn{3}{c}{\textbf{Consolidated Scores}} \\
\toprule
  \textbf{Model} & \textbf{macro F1 score}  & \textbf{weighted F1 score} \\
\toprule
ResNet 3D & 0.9095 & 0.9269\\
\hline
Pre-trained ResNet 3D & 0.8925 & 0.9171 \\
\hline
ResNet(2+1)D & 0.9035 & 0.9220 \\
\hline
Pre-trained ResNet(2+1)D  & 0.9237 & 0.9393 \\
\hline
ResNet Mixed Convolution &  0.8607 & 0.8881\\
\hline
\textbf{Pre-trained ResNet Mixed Convolution}\textcolor{orange}{*} & \textbf{0.9345} & \textbf{0.9470} \\

\toprule

\end{tabular}

\caption{Consolidated comparison of the models (\textcolor{orange}{*} denotes the overall winning model)}
\label{tab:f1_consolidated}
\end{table}%

ResNet Mixed Convolution with pre-training came up as the best model for both classes with pathology (LGG and HGG) and achieved a similar score as the other models while classifying healthy brain MRIs, as well as based on macro and weighted F1 scores - making this model as the clear overall winner. It can also be observed that the spatiospatial models performed better with pre-training, but ResNet 3D performed better without pre-training. 

\subsection{Comparison against literature}

\begin{table}[h!]
        \small
    \centering

    \begin{tabularx}{\linewidth}{@{} >{\RaggedRight\hsize=0.8\hsize}X
                                *{4}{>{\RaggedRight\hsize=1.05\hsize}X} @{}}
        \toprule
        Study & Method & Contrast & Dimension & Test Accuracy      \\ \addlinespace
        \midrule
    Shahzadi et al.~\cite{shahzadi2018cnn} &  CNN with LSTM & T2-FLAIR & 3D & 84.00 \%
    
 \\ \addlinespace 
        
        \\ \addlinespace
    
    Pei et al.~\cite{pei2019brain} & Similar to U-Net for Segmentation, Regular CNN for classification  & T1, T1ce, T2, T2-FLAIR & 3D & 74.9\% 
 \\ \addlinespace 
        
        \\ \addlinespace
    Ge et al.~\cite{ge2018deep} & Deep CNN & T1, T2 , T2-FLAIR &  2D & 90.87\% 
 \\ \addlinespace 
        
        \\ \addlinespace

     Yang et al.~\cite{yang2018glioma}
        & Pre-trained GoogLeNet    & T1ce & 2D   &  94.5\%\textsuperscript{\textdagger}         \\ \addlinespace 
        
        \\ \addlinespace 
        
     Mzoughi et al.~\cite{mzoughi2020deep}
        & Deep CNN       & T1ce & 3D &   96.49\%             \\   \addlinespace 
      
        \\ \addlinespace
    
     Zhuge et al.~\cite{zhuge2020automated}
        & Deep CNN  & T1, T1ce, T2, T2-FLAIR   & 3D &  97.1\% s*=0.968    \\ \addlinespace
        
         \\ \addlinespace
         
                 \\ \addlinespace
                 
     Ouerghi et al.~\cite{ouerghi2021glioma}\textsuperscript{\textdaggerdbl}
        & Random forest  & T1, T2, T2-FLAIR   & 2D &  96.5\%   \\ \addlinespace
        
         \\ \addlinespace
         
    \textbf{This paper} & \textbf{Pre-trained ResNet Mixed Convolution spatiospatial model} & \textbf{T1ce} & \textbf{3D} &   \textbf{96.98\%\textsuperscript{\textdagger}  s=0.9684}  \\

 \bottomrule
    \end{tabularx}
    \caption{Comparisons against other published works} 
    *s=specificity \textdagger cross-validated  \textdaggerdbl State of the art
    
    \label{tab:compare_sota}
    
    \end{table}

This sub-section compares the best model from the previous sub-section (i.e. ResNet Mixed Convolution with pre-training) against seven other research papers (in no specific order), where they classified LGG and HGG tumours. Mean test accuracy was used as the metric to compare the results as that was the common metric used in those papers. 

Starting from Shahzadi et al.~\cite{shahzadi2018cnn}, where they used LSTM-CNN to classify between HGG and LGG, using T2-FLAIR images from the BraTS 2015 dataset. Their work focuses on using a smaller sample size, and they were able to achieve 84.00\% accuracy~\cite{shahzadi2018cnn}. Pei et al.~\cite{pei2019brain} achieved even less classification accuracy of 74.9\% although they did use all of the available contrasts of the BraTS dataset, and their method performed segmentation using a U-Net like model before performing classification. Ge et al.~\cite{ge2018deep} uses a novel method of fusing the contrasts into multiple streams to be trained simultaneously. Their model achieved an accuracy of 90.87\% overall on all the contrasts, and they achieved 83.73\% on T1ce. Mzoughi et al.~\cite{mzoughi2020deep} achieved 96.59\% using deep convolutional neural networks on T1ce images. Their work does not present any other metric for their results, except for the overall accuracy of their model, which makes it difficult to compare against their results. Next, Yang et al.~\cite{yang2018glioma} did similar work; they used pre-trained GoogLeNet on 2D images, achieving an overall accuracy of 94.5\%. They did not use the BraTS dataset, but the purpose of their work was similar - to classify glioma tumours based on LGG and HGG grading. Their dataset had fewer samples of LGG and HGG class in comparison to this research, with the former having 52 samples, and later 61 samples~\cite{yang2018glioma}. Ouerghi et al.~\cite{ouerghi2021glioma} used different machine learning methods in their paper to train on the fusion images, one of which is the random forest, on which they achieved 96.5\% for classification between High-Grade and Low-Grade Glioma. Finally, the Zhuge et al.~\cite{zhuge2020automated} achieved an impressive 97.1\% using Deep CNN for classification of glioma based on LGG and HGG grading, beating the proposed model by 0.12\%. This difference can be explained by two factors, 1) their use of an additional dataset from The Cancer Imaging Archive (TCIA) in combination with BraTS 2018 2) and their use of four different contrasts - both these factors increase the size of the training set significantly. Furthermore, no cross-validation has been reported in their paper. Table~\ref{tab:compare_sota} shows the complete comparative results. 


\section{Discussion}
The F1 scores of all the models in classifying healthy brains were very close to one, as segregating healthy brains from brains with pathology is comparatively a simpler task than classifying the grade of the tumour. Furthermore, using two different datasets for healthy and pathological brain, MRIs could have also introduced a dataset bias. In classifying the grade of the tumour, the pre-trained ResNet Mixed Convolution model performed best, while in classifying healthy brains, all the three pre-trained models performed similarly. For comparing the models based on consolidated scores, macro and weighted F1 scores were used. However, the macro F1 score is to be given more importance as the dataset was imbalanced. Both of the metrics declared the pre-trained ResNet Mixed Convolution as the clear winner.

One interesting observation that can be made from the confusion matrices is that the classification performance of the models for the LGG class has been lower than the other two classes. Even the best performing model managed to get an accuracy of 81\% for LGG while achieving 96\% for HGG and nearly perfect results for healthy. This might be attributed to the fact that the dataset was highly imbalanced (Sec.~\ref{sec:dataset}), i.e. 259 volumes each for HGG and healthy, while having 73 volumes for LGG. Even though weighted cross-entropy loss (Sec.~\ref{sec:loss}) was used in this research to deal with the problem of class imbalance, increasing the number of LGG samples or employing further techniques to deal with this problem further and might improve the performance of the models for LGG~\cite{johnson2019survey}. 

It is noteworthy that the pre-trained ResNet Mixed Convolution resulted in the best classification performance, even though it is the model with the least number of trainable parameters (see Table~\ref{tab:params}). Moreover, it is to be noted that both spatiospatial models performed better than the pure 3D ResNet18 model, even though they had a fewer number of trainable parameters than the 3D ResNet18. A fewer number of trainable parameters can reduce the computational costs, as well as the chance of overfitting. The authors hypothesise that the increased non-linearity due to the additional activation functions between the 2D and 1D convolutions in (2+1)D convolutional layers helped the ResNet (2+1)D model to achieve better results than ResNet3D, and the reduction of trainable parameters while having a similar number of layers, in turn preserving the level of non-linearity, contributed to the success of ResNet Mixed Convolution. Even though it has been seen that the spatiospatial models performed better, it is worthy of mention that the spatiospatial models do not adequately maintain the 3D nature of the data - the spatial relationship between the three dimensions is not preserved within the network like a fully 3D network as ResNet3D - which is a limitation of this architecture, which might have some unforeseen adverse effects. The authors hypothesised that this relationship was indirectly maintained through the channels of the network, and the network could learn the general representation to be able to classify appropriately. The experiments have also shown that the spatiospatial models are superior to a fully 3D model for the brain tumour classification problem shown here. Nevertheless, before creating a common consensus about this finding, these models should be further evaluated for other tasks.

In this research, the slice dimension in the axial orientation was considered as the "specially-treated" spatial dimension of the spatiospatial models, which can also be seen as the pseudo-temporal dimension of the spatiotemporal models. The authors hypothesise that using the data in sagittal or coronal orientation in a similar way might also be possible to exploit the advantages of such models, which it is yet to be tested.

It can also be observed that the pre-trained models were the winners for all three different classes. However, the effect of pre-training was not the same on all three models. For both the spatiospatial models, pre-training improved the model's performance, but in different amounts: $2.24\%$ improvement for ResNet (2+1)D and $8.57\%$ for ResNet Mixed Convolution (based on macro F1 scores). However, pre-training had a negative impact on the 3D ResNet18 model (for two out of three classes), causing it to decrease the macro F1 score by $1.87\%$. Average macro F1 scores for all the models with and without pre-training (0.9169 with pre-training, 0.8912 without pre-training) show that the pre-training resulted in an overall improvement of $2.88\%$ across models. It is noteworthy that the pre-trained networks were initially trained on RGB videos. Pre-training them on MRI volumes or MR videos (dynamic MRIs) might further improve the performance of the models.

Regarding the comparisons to other published works, an interesting point to note is that the previous papers only classified different grades of brain tumours (LGG and HGG), whereas this paper also classified healthy brains as an additional class. Thus, the results are not fully comparable as more classes increase the difficulty of the task. Even then, the results obtained by the winning model are better than all previously published methods, except for one, which reported comparable results to the ResNet Mixed Convolution (that paper reported $0.12\%$ better accuracy, and $0.41\%$ less specificity). However, this paper used four different contrasts and an additional dataset apart from BraTS, making them have a larger dataset for training. 

\section{Conclusion}
This paper shows that the spatiotemporal models, ResNet(2+1)D and ResNet Mixed Convolution, working as spatiospatial models, could improve the classification of grades of brain tumours (i.e. low-grade and high-grade glioma), as well as classifying brain images with and without tumours, while reducing the computational costs. A 3D ResNet18 model was used to compare the performance of the spatiospatial models against a pure 3D convolution model. Each of the three models was trained from scratch and also trained using weights from pre-trained models that were trained on an action recognition dataset - to compare the effectiveness of pre-training in this setup. The final results were generated using cross-validation with three folds. It was observed that the spatiospatial models performed better than a pure 3D convolutional ResNet18 model, even though having fewer trainable parameters. It can be observed further that pre-training improved the performance of the models. Overall, the pre-trained ResNet Mixed Convolution model was observed to be the best model in terms of F1-score, obtaining a macro F1-score of 0.9345 and a mean test accuracy of 96.98\%, while achieving 0.8949 and 0.9123 F1-scores for low-grade glioma and high-grade glioma, respectively. This study shows that the spatiospatial models have the potential to outperform a fully 3D convolutional model. However, this was only shown for a specific task here - brain tumour classification, using one dataset - BraTS. These models should be compared for other tasks in the future to build a common consensus regarding the spatiospatial models. One limitation of this study is that it only used T1 contrast-enhanced images for classifying the tumours, which already resulted in good accuracy. Incorporating all four available types of images (T1, T1ce, T2, T2-Flair) or any combination of them might improve the performance of the model even further.

\section*{Acknowledgement}

This work was in part conducted within the context of the International Graduate School MEMoRIAL at Otto von Guericke University (OVGU) Magdeburg, Germany, kindly supported by the European Structural and Investment Funds (ESF) under the programme "Sachsen-Anhalt WISSENSCHAFT Internationalisierung" (project no. ZS/2016/08/80646).

\bibliography{bibliography}

\end{document}